\documentclass[preprint,prd,aps,nofootinbib]{revtex4}
\usepackage{amsmath}
\usepackage{graphicx}
\usepackage{subcaption}
\usepackage{tikz}
\usepackage{diagbox}
\usepackage{slashed}
\usepackage{float}
\usepackage{color}
\usepackage{makecell}
\begin{document}
\title{$S-P-D$ Mixing in Vector Quarkonia from the Salpeter Equation with Optimized Wave Function Representations}
\author{Wen-Yuan Ke$^{1,2,3,4}$,
Qiang Li$^5$, Tianhong Wang$^6$, Tai-Fu Feng$^{1,2,3,4}$,
Guo-Li Wang$^{1,3,4}$\footnote{corresponding author}}
\affiliation{$^1$ Department of Physics, Hebei University, Baoding, 071002, China\\
$^2$Department of Physics, Guangxi University, Nanning, 530004, China\\
$^3$  Hebei Key Laboratory of High-Precision Computation and Application of Quantum Field Theory, Baoding 071002, China\\
$^4$ Hebei Research Center of the Basic Discipline for Computational Physics, Baoding 071002, China\\
$^5$ School of Physical Science and Technology, Northwestern Polytechnical University, Xi'an 710072, China\\
$^6$ School of Physics, Harbin Institute of Technology, Harbin 150001, China
}

\begin{abstract}
This paper proposes a novel mechanism based on the instantaneous Bethe-Salpeter (Salpeter) equation for investigating wave function mixing in vector mesons such as $\psi(3770)$. Conventional theories typically treat $\psi(3770)$ as a $2S-1D$ mixed state; however, considering only tensor forces or relativistic corrections alone often leads to mixing angles that are too small and inconsistent with experimental data. Phenomenological $2S-1D$ mixing requires experimental data as input to determine the mixing angles, resulting in limited theoretical studies on states like $\Upsilon(1D, 2D)$ in the absence of experimental data. To more accurately describe $S-D$ mixing and its relativistic effects, this paper systematically compares four relativistic wave function representations ($\varphi_1$, $\varphi_2$, $\varphi_3$, and $\varphi_4$) by solving the Salpeter equation and calculates the mass spectra and dileptonic decay widths of charmonium and bottomonium. The study finds that the wave function representation $\varphi_2$ can simultaneously reproduce the experimental data of both charmonium and bottomonium well. Further analysis reveals that, in addition to $S-D$ mixing, the wave functions of vector mesons contain a non-negligible $P$-wave component, meaning they are $S-P-D$ mixed states. We predict the mixing angles for bottomonium $\Upsilon(1D)$ and $\Upsilon(2D)$ to be $(1.78^{+0.32}_{-0.25})^\circ$ and $(5.44^{+1.10}_{-0.76})^\circ$, with dileptonic decay widths of $2.29^{+0.86}_{-0.69}$ eV and $10.5^{+4.2}_{-3.1}$ eV, respectively.

\end{abstract}

\maketitle

\section{Introduction}

The $\psi(3770)$ is a vector charmonium primarily dominated by the $D$‑wave with a small $S$‑wave admixture. It was experimentally discovered in 1977 \cite{early0}. However, before its discovery, theoretical predictions had indicated that the tensor force between quarks, which does not conserve orbital angular momentum, would lead to $S-D$ mixing \cite{early1}. Beside the tensor‑force, coupled‑channel effects can also account for the $S-D$ mixing \cite{early5,rela2}. Consequently, $1^-$ mesons with the same quark composition all have the potential to mix, including highly excited states \cite{rela2,Moremix1,Moremix2,maozhi}.
Since the masses of the $\psi(2S)$ and $\psi(3770)$ are close, their mixing is maximal \cite{early1}.
It is therefore widely accepted that $\psi(2S)$ and $\psi(3770)$ are $2S-1D$ mixing states. The mixing formula is
\begin{align}\label{mix}
|\psi(2S)\rangle = \cos\theta |{2}^{3}S_1\rangle - \sin\theta |{1}^{3}D_1\rangle,\nonumber \\
|\psi(3770)\rangle = \sin\theta |{2}^{3}S_1\rangle + \cos\theta |{1}^{3}D_1\rangle .
\end{align}

The $S-D$ mixing nature of $\psi(3770)$ has been confirmed experimentally. The decay rate of a pure $D$‑wave state into dileptons is extremely small, vanishing in the non‑relativistic limit. The admixture of an $S$ ‑wave component significantly enhances the decay width, making it experimentally observable \cite{early0}.
Subsequently, the mixing properties of $\psi(3770)$ have attracted substantial research interest \cite{early3,Eichten:2007qx}.
Further study finds that the naive picture of $\psi(3770)$ as a simple $1D-2S$ mixed state is unsatisfactory. Many theories predict a mixing angle around $|\theta| =10^\circ$; for example, Ref. \cite{Ding:1991vu} gives $10^\circ$, Ref. \cite{smallangle1} gives $12^\circ$, and Ref. \cite{smallangle2} gives $(12\pm 2)^\circ$, among others.
Nevertheless, both smaller angles such as $5.4^\circ$ \cite{smallangle3} and much larger angles such as $(17.4\pm 2.5)^\circ$ \cite{Barnes:2005pb} and  $40^\circ$ \cite{Liu:2004un} have also been reported.

Furthermore, Ref. \cite{Voloshin} points out that the non‑zero contribution of $\psi(3770)$ to the dilepton arises from two mechanisms. The first is the $S-D$ mixing induced by the tensor force, and the second originates from relativistic corrections. These two mechanisms interfere and are of the same order at $v^2$. Therefore, it is generally unreasonable to consider only one mechanism while neglecting the other. Ref. \cite{Moremix3} also indicates that the tensor force is too weak to yield a sufficiently large mixing angle, further implying that relativistic corrections must be included to explain the $S-D$ mixing of $\psi(3770)$ \cite{rela1,rela2}.
Moreover, 
there is also the spin‑orbit interaction  in the potential, which violates total spin conservation and mixes different total spins. Since total angular momentum is conserved, this can result in, for example, $S-P$ ($P-D$) mixing, a phenomenon that has not yet been studied in detail \cite{peisy1}.

Similar to charmonium, $S-D$ mixing must also occur in bottomonium. However, Ref. \cite{zhong} points out that theoretical studies often overlook the underlying dynamic mechanisms responsible for $S-D$ mixing and artificially introduce mixing angles. This necessitates determining mixing angles by fitting experimental data, making it difficult to study $\Upsilon(1D)$ and $\Upsilon(2D)$, for which no experimental data currently exist. Consequently, theoretical studies on their mixing are extremely limited. Therefore, it is necessary to conduct a thorough investigation into their mixings to facilitate their experimental discovery.

We note that in traditional approaches, relativistic potentials are derived by studying quark‑antiquark scattering  \cite{Gupta}, where the quark spinors are expanded into non‑relativistic Pauli spinors, and all quantities except the Pauli spinors are absorbed into the interaction potential, thereby obtaining a relativistic potential. In this method, the potential is relativistic, while the wave function of bound state remains non‑relativistic  \cite{Gupta}. We adopt an opposite approach: the wave function is relativistic, while the potential is non‑relativistic \cite{Wang:2005qx,ee}.
The advantage of the former is that it yields a relativistic mass spectrum, while in our approach, since we solve the instantaneous Bethe‑Salpeter equation \cite{Salpeter}, both the mass spectrum and the wave function are relativistic.

The Bethe-Salpeter (BS) equation is a relativistic dynamical equation describing bound states \cite{BS equation}. However, similar to the Schr\"{o}dinger equation, the form of the wave function is externally input. Previously, for the $1^-$ vector mesons, we proposed a general wave function representation in the instantaneous approximation \cite{Wang:2005qx,ee}. Yet this is not the only or inevitable representation. In this paper, we will present several possible relativistic wave function representations, substitute them into the BS equation for solving, and determine the optimal wave function by considering the mass spectrum and calculating annihilation decays.
As an iterative integral equation, the BS equation incorporates tensor forces and relativistic corrections to infinite orders.  The wave function also accounts for the mixing of different partial waves, $S-P-D$ mixing instead of only $S-D$ mixing \cite{peisy1}.

\section{Introduction of the Salpeter equation}

The BS equation is difficult to solve exactly, so we rigorously solve its instantaneous approximation, the Salpeter equation. The instantaneous approximation manifests as the interaction being independent of the time component, namely,
$V(P,k,q)\sim V(\vec{k},\vec{q})=V(\vec{q}-\vec{k})$, where $P$ is the total momentum of the meson, $q$ ($k$) is the relative momentum between the two quarks inside the meson.

Define the positive and negative energy projection operators as
$\Lambda_{i}^{\pm}(q_{\perp}) = \frac{1}{2 \omega_{i }}\left[\frac{\slashed P}{M} \omega_{i } \pm J(i)(m_{i}+\slashed q_{{\perp}})\right]$,
where $J(i) = (-1)^{i+1}$, $i=1$ for quark, and $i=2$ for antiquark;  $M$ and $m_i$ are the masses of meson and inside quark, respectively; $\omega_{i} = \sqrt{m_{i}^{2}-q_{{\bot}}^2}$ is the energy of quark. We have defined $q_{{\bot}}\equiv q-\frac{P\cdot q}{M^2}P$, so $q_{{\bot}}=(0,\vec{q})$ in the center-of-mass system of $P$.
Applying definition,
$
\varphi^{\pm\pm}(q_{\perp}) \equiv \Lambda_{1}^{\pm}(q_{\perp}) \frac{\slashed P}{M} \varphi(q_{\perp}) \frac{\slashed P}{M} \Lambda_{2 }^{\pm}(q_{\perp}),
$
the wave function is decomposed into four terms
$
\varphi = \varphi^{++} + \varphi^{+-} + \varphi^{-+} + \varphi^{--}
$. With these notations, the Salpeter equation is written as \cite{Salpeter},
\begin{align} \label{Lorentz-covariant}
(M - \omega_{1} - \omega_{2}) \varphi^{++}(q_{\perp}) &= \Lambda_{1}^{+}(q_{\perp}) \left[ \int \frac{d\vec{k}}{(2 \pi)^{3}} V({q}_{\perp}-k_{\perp}) \varphi({k}_{\perp})\right] \Lambda_{2}^{+}(q_{\perp}), \nonumber\\
(M + \omega_{1} + \omega_{2}) \varphi^{--}(q_{\perp}) &= -\Lambda_{1 }^{-}(q_{\perp})\left[ \int \frac{d\vec{k}}{(2 \pi)^{3}} V({q}_{\perp}-k_{\perp}) \varphi({k}_{\perp})\right] \Lambda_{2}^{-}(q_{\perp}), \\
\varphi^{+-}(q_{\perp}) &= \varphi^{-+}(q_{\perp}) = 0. \nonumber
\end{align}
The corresponding normalization condition is
\begin{equation}\label{normalization condition}
\int \frac{d\vec{q}}{(2 \pi)^{3}} \operatorname{Tr}\left[\bar{\varphi}^{++}(q_{\perp}) \frac{\slashed P}{M} \varphi^{++}(q_{\perp}) \frac{\slashed P}{M} - \bar{\varphi}^{--} (q_{\perp}) \frac{\slashed P}{M} \varphi^{--}(q_{\perp}) \frac{\slashed P}{M}\right] = 2 M,
\end{equation}
where, $\bar{\varphi}=\gamma_0 \varphi^{\dag}\gamma^0$, `$\dag$' is the Hermitian conjugate transformation.

Since the wave functions are relativistic, to avoid double counting, we must choose a non-relativistic interaction integral kernel. We adopt the modified Cornell potential \cite{Wang:2005qx},
\begin{eqnarray}\label{potential2}
V(\vec{q}) = -(\frac{\lambda}{\alpha}+V_{0})\delta^{3}(\vec{q})+\frac{\lambda}{\pi^{2}}\frac{1}{(\vec{q}^{2}+\alpha^{2})^{2}}
-\gamma_{0}\otimes\gamma^{0}\frac{2}{3\pi^{2}}\frac{\alpha_{s}(\vec{q})}{(\vec{q}^{2}+\alpha^{2})},
\end{eqnarray}
where $\lambda$ is the string tension, $V_0$ is a free parameter, $\alpha$ is a small quantity to avoid infrared divergence and account for screening effects, and $\alpha_{s}(\vec{q})=\frac{4\pi}{9}\frac{1}{log(e+{\vec{q}^2}/{\Lambda_{QCD}^{2}})}$ is running coupling constant, $\Lambda_{QCD}$ is the QCD scale and $e=2.7183$.

\section{Choice of Wave Function representations}

Similar to solving the Schr\"{o}dinger equation, when solving the Salpeter equation, it is essential to first specify a concrete representation of the wave function.
We know that the orbital angular momentum $l$ is not always a good quantum number, as can be seen from the fact that the particle $\psi(3770)$ is an $S-D$ wave mixing state. In contrast, the total angular momentum $J$ is always a good quantum number; hence, we express the relativistic wave function of a particle in terms of its total angular momentum, specifically, in terms of its $J^P$ quantum number ($P$ is the parity).
In quantum field theory, the general relativistic wave function of a vector meson with quantum number $J^P=1^-$, constructed via Dirac matrices, contains 16 independent terms. However, under the instantaneous approximation, the 8 terms containing $P\cdot q\equiv P\cdot q_{\perp}=0$ vanish. Consequently, the general wave function of a $1^-$ meson consists of 8 terms \cite{Wang:2005qx}:
\begin{align}\label{1-}
\varphi_{1^{-}}^{\text{full}} = \epsilon \cdot {q_{}}_{\perp} \left( f_1 +  \frac{\slashed{P}}{M}f_2+ \frac{\slashed{q}_{\perp}}{M} f_3 + \frac{\slashed{P}\slashed{q}_{\perp}}{M^2} f_4 \right) + \left(M f_5 + \slashed{P} f_6\right)\slashed{\epsilon}+\left(  f_7 + {\frac{\slashed{P}}{M} f_8}\right){\slashed{\epsilon} \slashed{q}_\perp}.
\end{align}
The radial wave functions $f_i \equiv f_i(-q^2_{\perp})~(i=1,2,...,8)$ are functions of $q^2_{\perp}$.
Next, we demonstrate that each term in the proposed $1^-$ wave function possesses negative parity. When we perform parity transformation, $P'=(P_0,-\vec{P})$ , $q'=(q_0,-\vec{q})$. While in the condition of the center-of-mass system of $P$ and the instantaneous approximation, we have  $P'=P$ and $q_{\perp}'=-q_{\perp}$. Under parity transformation, the wave functions satisfy the relation:
$$
\varphi^{\rm full}_{1^{-}}\left(P, q_{\perp}\right)=\eta_{_P} \gamma_{0} \varphi^{\rm full}_{1^{-}}\left(P', q_{\perp}'\right) \gamma_{0},
$$
where $\eta_{_P}$ is the parity, and
$$
\begin{aligned}
\gamma_{0} \varphi^{\rm full}_{1^-}(P',q_\perp') \gamma_0  = -\epsilon_{\mu} {q_\perp^\mu}\gamma_0  \left(f_1 + \frac{\slashed P}{M} f_2 - \frac{\slashed q_\perp}{M} f_3- \frac{\slashed P \slashed q_\perp}{M^2} f_4 \right) \gamma_0  \\
 + \gamma_0 \left(M f_5+\slashed P f_6\right)\slashed\epsilon  \gamma_0 -\gamma_0 \left( f_7+\frac{\slashed P}{M} f_8\right) \slashed\epsilon\slashed q_\perp\gamma_0= -\varphi^{\rm full}_{1^{-}}(P, q_\perp).
\end{aligned}
$$
Therefore  $\eta_{_P}=-1$.  Similarly, by applying the charge conjugation transformation
$$
\varphi^{\rm full}_{1^{-}}\left(P, q_{\perp}\right)=\eta_{_C} C \left[\varphi^{\rm full}_{1^{-}}\left(P', q_{\perp}'\right)\right]^T C^{-1},
$$ where $C$ is the charge conjugate transform, $T$ is the rotation transform, $C\gamma^T_{\mu}C^{-1}=-\gamma_{\mu}$, and $\eta_{_C}$ is the charge conjugate parity. We find that the $f_2$ and $f_7$ terms have positive charge conjugation parity, while the remaining terms have negative parity. However, our theory is self-consistent; in the case of quarkonium, since $\varphi^{+-}=\varphi^{-+}=0$ and $m_1=m_2$, we have $f_2=f_7=0$ and then $\eta_{_C}=-1$. Therefore, for quarkonium, the wave function of Eq. (\ref{1-}) has $J^{PC}=1^{--}$.

We note that the terms containing $f_5$ and $f_6$ in Eq. (\ref{1-}) are $S$-waves with angular momentum $l=0$.  While the $f_1$, $f_2$, $f_7$ and $f_8$ terms are $P$-waves with $l=1$, they have negative parity and do not satisfy the formula $P=(-1)^{l+1}$. In Ref. \cite{wangBc}, we have pointed out that the parity formula $P=(-1)^{l+1}$ holds only when $l$ is a good quantum number, or equivalently, in the non-relativistic case.
The $f_3$ and $f_4$  terms include both $S$-waves and $D$-waves. Since $(\epsilon\cdot q_{_\bot})\slashed{q}_{\bot}=\frac{1}{3}q_{_\bot}^{2}\slashed{\epsilon}+\left[(\epsilon\cdot q_{_\bot})\slashed{q}_{\bot}-\frac{1}{3}q_{_\bot}^{2}\slashed{\epsilon}\right]$, where $\frac{1}{3}q_{_\bot}^{2}\slashed{\epsilon}$ is the $S$-wave and $(\epsilon\cdot q_{_\bot})\slashed{q}_{\bot}-\frac{1}{3}q_{_\bot}^{2}\slashed{\epsilon}$ is the $D$-wave \cite{peisy1}.

{We emphasize that the $P$-wave components identified here, namely the $f_1$, $f_2$, $f_7$, and $f_8$ terms, are relativistic components arising from the Dirac-covariant decomposition in Eq.~(\ref{1-}),
and they cannot represent an independently conserved nonrelativistic $P$-wave state. As we have shown earlier, they carry negative parity, in contrast to the positive parity possessed by a $P$-wave state. Moreover, with different quantum number, their representations also differ from those of nonrelativistic $P$-wave states. For example, the nonrelativistic wave-function representation of a $P$-wave $1^+$ state is given by $\varphi_{_{1^+}}=g^{}_1\epsilon \cdot q^{}_{_\bot}\left(1+{\not\!P}/{M}\right)\gamma^{5}
+{ig^{}_2(1+{\not\!P}/{M})\varepsilon^{}_{\mu\nu\rho\sigma}
\gamma^{\mu}_{}P^{\nu}_{}q^{\rho}_{_\bot}\epsilon^{\sigma}}/M$ \cite{wangBc}, where $g^{}_1$ and $g^{}_2$ are functions of $-q^{2}_{_\bot}$. It is evident that the Lorentz structure of the nonrelativistic wave function for a $P$-wave state is clearly different from the ones of the $P$-wave components $f_1$, $f_2$, $f_7$, and $f_8$ for the $1^-$ state given here.}

In Eq. (\ref{1-}), the radial wave functions $f_1$, $f_2$,..., $f_8$ are unknown, and their numerical solutions are obtained by solving the dynamical equation, namely the Salpeter equation. After solving, we find that for any $1^-$ meson system, the states corresponding to the solutions appear in the following order: $1S$, $2S$, $1D$, $3S$, $2D$, ... Of course, none of these wave functions are pure; for instance, the $nS$ state wave function is predominantly $S$-wave, with a small admixture of $P$-wave and an even smaller $D$-wave component (we will show this later).
Using the wave function in Eq. (\ref{1-}), we studied various $1^-$ mesons and found that for $S$-wave-dominated mesons such as $D^*$, $D_s^*$, $J/\psi$, and their radial excited states, our theoretical results align well with experimental data \cite{BtoD,BstoDs,peisy,jiam,ee}. However, for $D$-wave-dominated mesons like $\psi(3770)$, although their masses and strong decay results match experiments well \cite{massspa}, the annihilation results to dileptons deviates from experimental data. This indicates an unreasonable $S-D$ wave mixing in the wave function, suggesting that the expression Eq. (\ref{1-}) requires improvement.

Specifically, when constructing a wave function representation that satisfies the $1^-$ quantum numbers, it may not be necessary to include all terms.
For mesons dominated by $S$-waves, the $f_5$ and $f_6$ terms are necessary, while for those dominated by $D$-waves, the $f_3$ and $f_4$ terms are required. As for a meson with $S-D$ mixing, all four terms must be included. As for the proportion of each partial wave (including $P$-wave) within the wave function, it is determined by the dynamic equation they satisfy, namely the Salpeter equation.
Therefore, we present the following four relativistic wave function representations for the $1^{--}$ quarkonium,
\begin{align} \label{wave}
& (1)~\varphi_{\text{1}}= \epsilon \cdot {q_{}}_{\perp} \left( f_1 + \frac{\slashed{q}_{\perp}}{M} f_3 + \frac{\slashed{P}\slashed{q}_{\perp}}{M^2} f_4 \right) +  \left(M f_5 + \slashed{P} f_6\right)\slashed{\epsilon} + {\frac{\slashed{P}\slashed{\epsilon} \slashed{q}_\perp}{M} f_8},  \nonumber  \\
& (2)~\varphi_{\text{2}}= \epsilon \cdot {q_{}}_{\perp} \left( f_1 + \frac{\slashed{q}_{\perp}}{M} f_3 + \frac{\slashed{P}\slashed{q}_{\perp}}{M^2} f_4 \right) + \left(M f_5 + \slashed{P} f_6\right)\slashed{\epsilon}, \nonumber \\
& (3)~\varphi_{\text{3}}= \epsilon \cdot {q_{}}_{\perp} \left(\frac{\slashed{q}_{\perp}}{M} f_3 + \frac{\slashed{P}\slashed{q}_{\perp}}{M^2} f_4 \right) +  \left(M f_5 + \slashed{P} f_6\right)\slashed{\epsilon}+ {\frac{\slashed{P}\slashed{\epsilon} \slashed{q}_\perp}{M} f_8}, \nonumber \\
& (4)~\varphi_{\text{4}}= \epsilon \cdot {q_{}}_{\perp} \left( f_1 + \frac{\slashed{q}_{\perp}}{M} f_3 + \frac{\slashed{P}\slashed{q}_{\perp}}{M^2} f_4 \right) + {\frac{\slashed{P}\slashed{\epsilon} \slashed{q}_\perp}{M} f_8}.
\end{align}
where $\varphi_{\text{1}}$ is the general wave function representation of the $1^{--}$ state, i.e., the one appears in Eq. (\ref{1-}).

For the wave functions of a $1^{--}$ state, except the expressions in Eq. (\ref{wave}), there are another possibilities.
For example, we have tried the wave functions $\left(M + \slashed{P}\right)\slashed{\epsilon}f_5$ and $\left(M f_5 + \slashed{P} f_6 \right)\slashed{\epsilon}$, and found that their corresponding solutions are the $1S$, $2S$, $3S$, ... states, with the wave functions containing only $S$-waves.
We also tried $(\epsilon\cdot q_ {\perp}-\frac{1}{3} \slashed {\epsilon}\slashed {q} _ {\perp}) \left( \frac{\slashed{q}_{\perp}}{M} + \frac{\slashed{P}\slashed{q}_{\perp}}{M^2} \right) f_3$, $(\epsilon\cdot q_ {\perp}-\frac{1}{3} \slashed {\epsilon}\slashed {q} _ {\perp}) \left( \frac{\slashed{q}_{\perp}}{M} f_3 + \frac{\slashed{P}\slashed{q}_{\perp}}{M^2} f_4 \right)$, $\epsilon \cdot {q}_{\perp} \left( \frac{\slashed{q}_{\perp}}{M} + \frac{\slashed{P}\slashed{q}_{\perp}}{M^2} \right) f_3$ and $\epsilon \cdot {q}_{\perp} \left( \frac{\slashed{q}_{\perp}}{M} f_3 + \frac{\slashed{P}\slashed{q}_{\perp}}{M^2} f_4 \right)$, and found that their solutions correspond to the $1D$, $2D$, $3D$, ... states, where the first two wave functions are purely $D$-waves, while the last two contain both $D$- and $S$-waves but is dominated by the $D$-wave. The above wave functions can all exist individually as solutions to the Salpeter equation, but we consider them to be non-relativistic wave functions, not the relativistic case we are concerned with, and therefore they are therefore omitted in Eq. (\ref{wave}).

Additionally, we also attempted the wave functions $\epsilon \cdot {q_{}}_{\perp} \left( f_1 +  \frac{\slashed{P}}{M}f_2 \right)$, $\left(  f_7 + {\frac{\slashed{P}}{M} f_8}\right){\slashed{\epsilon} \slashed{q}_\perp}$, $\epsilon \cdot {q_{}}_{\perp} \left( f_1 +  \frac{\slashed{P}}{M}f_2 \right) +\left(  f_7 + {\frac{\slashed{P}}{M} f_8}\right){\slashed{\epsilon} \slashed{q}_\perp}$, $\epsilon\cdot q_ {\perp}\left( \frac{\slashed{q}_{\perp}}{M} f_3 + \frac{\slashed{P}\slashed{q}_{\perp}}{M^2} f_4 \right) + \left(M f_5 + \slashed{P} f_6\right)\slashed{\epsilon}$, and $(\epsilon\cdot q_ {\perp}-\frac{1}{3} \slashed {\epsilon}\slashed {q} _ {\perp}) \left( \frac{\slashed{q}_{\perp}}{M} f_3 + \frac{\slashed{P}\slashed{q}_{\perp}}{M^2} f_4 \right) + \left(M f_5 + \slashed{P} f_6\right)\slashed{\epsilon}$,
but when applying the last two constraint conditions $\varphi^{+-}=\varphi^{-+}=0$ in Eq. (\ref{Lorentz-covariant}), we found that the radial parts of these wave functions all vanish,
$f_1=f_2=f_7=f_8=0$, indicating that these wave function representations cannot exist independently as a solution to the Salpeter equation. We have also attempted to calculate the wave functions of the forms $\epsilon\cdot q_ {\perp} f_1 + (\epsilon\cdot q_ {\perp}-\frac{1}{3} \slashed {\epsilon}\slashed {q} _ {\perp}) \left( \frac{\slashed{q}_{\perp}}{M} f_3 + \frac{\slashed{P}\slashed{q}_{\perp}}{M^2} f_4 \right) + \left(M f_5 + \slashed{P} f_6\right)\slashed{\epsilon} + {\frac{\slashed{P}\slashed{\epsilon} \slashed{q}_\perp}{M} f_8}$, $\epsilon\cdot q_ {\perp} f_1 + (\epsilon\cdot q_ {\perp}-\frac{1}{3} \slashed {\epsilon}\slashed {q} _ {\perp}) \left( \frac{\slashed{q}_{\perp}}{M} f_3 + \frac{\slashed{P}\slashed{q}_{\perp}}{M^2} f_4 \right) + \left(M f_5 + \slashed{P} f_6\right)\slashed{\epsilon}$, and $(\epsilon\cdot q_ {\perp}-\frac{1}{3} \slashed {\epsilon}\slashed {q} _ {\perp}) \left( \frac{\slashed{q}_{\perp}}{M} f_3 + \frac{\slashed{P}\slashed{q}_{\perp}}{M^2} f_4 \right) +  \left(M f_5 + \slashed{P} f_6\right)\slashed{\epsilon} + {\frac{\slashed{P}\slashed{\epsilon} \slashed{q}_\perp}{M} f_8}$, and found that their results are identical to those of $\varphi_1$, $\varphi_2$, and $\varphi_3$ in Eq. (\ref{wave}), respectively. It is straightforward to prove that the former and the latter are equivalent. 

Although the wave function representation is artificially introduced, the Salpeter equation imposes constraints on it, the last two sub-equations of the Eq. (\ref{Lorentz-covariant}), $\varphi^{+-}(q_{\perp}) = \varphi^{-+}(q_{\perp}) = 0$, establish relations among the different radial wave functions, reducing the number of independent radial wave functions to four (two for $\varphi_{\text{4}}$). We choose the independent radial wave functions as $f_3$, $f_4$, $f_5$, and $f_6$ ($f_3$ and $f_4$ for $\varphi_{\text{4}}$), while the relations of the other radial wave functions to them are listed in Table \ref{rela}.
The table also provides the corresponding normalization formulas satisfied by the radial wave functions under different choices of wave function representation.

\begin{table}[ht]
\begin{center}
\caption{Relations and normalization formulas for radial wave functions with different wave function (WF) choices.}
\label{rela}
\setlength{\tabcolsep}{1.5mm}{
\begin{tabular}{c c c}
\hline
WF &  Relations  &  Normalization Formulas  \\
\hline
$\scriptstyle\varphi_{1}$ &\makecell[t]{ $ \scriptstyle f_1=\frac{ f_3 q_ {\perp}^2 + f_5 M^2}{M m_1},~\scriptstyle f_8 = \frac{f_6 M}{m_1}$}   & $\scriptstyle \int \frac{d\vec{q}}{(2\pi)^3} \frac{4Mw_1}{3m_1}\left[ 3f_5f_6+\frac{q_\perp^2}{M^2}\left(\frac{q_\perp^2}{M^2}f_3f_4+f_3f_6 +f_4f_5 \right) \right]=1$\\
$\scriptstyle\varphi_{2}$ & $ \scriptstyle f_1=\frac{ f_3 q_ {\perp}^2 + f_5 M^2}{M {m_1}}$   &
$\scriptstyle \int \frac{d\vec{q}}{(2\pi)^3} \frac{4Mw_1}{3m_1}\left[ f_5f_6\left(3+\frac{2q_\perp^2}{w_1^2}\right)+\frac{q_\perp^2}{M^2}\left(\frac{q_\perp^2}{M^2}f_3f_4 +f_3f_6 +f_4f_5 \right) \right]=1$ \\
$\scriptstyle\varphi_{3}$ & $ \scriptstyle f_8 = \frac{f_6 M}{m_1}$    & $\scriptstyle \int \frac{d\vec{q}}{(2\pi)^3} \frac{4Mm_1}{3w_1}\left[ f_5f_6\left(3-\frac{2q_\perp^2}{m_1^2}\right)+\frac{q_\perp^2}{M^2}\left(\frac{q_\perp^2}{M^2}f_3f_4+f_3f_6 +f_4f_5 \right) \right]=1$ \\
$\scriptstyle\varphi_{4}$ & $ \scriptstyle  f_1 = \frac{f_3 q_ {\perp}^2}{M m_1}, ~f_8=0$   & $\scriptstyle \int \frac{d\vec{q}}{(2\pi)^3} \frac{4 q_{\perp}^4 w_1 }{3 M^3 m_1}f_3 f_4 =1$ \\
\hline
\end{tabular}}
\end{center}
\end{table}

We substitute the wave function representation containing four independent unknown radial wave functions into the first two equations of Eq. (\ref{Lorentz-covariant}) for solving. Since both the equations and the wave functions contain Dirac matrices, we remove the matrices by taking the trace and find that the number of independent equations is not two on the surface but actually four. With four equations and four unknowns, the Salpeter equation can thus be solved successfully, yielding the numerical results for the radial wave functions and the mass spectrum.

The decay width of a vector quarkonium into a lepton-antilepton pair is proportional to the square of its decay constant, and the formula for calculating the decay constant  $F_V$ of a vector from its wave function is:
\begin{equation}
F_V M\epsilon_{\mu}=\sqrt{N_c}\int\frac{d^4q}{(2\pi)^4}\mathrm{Tr}[\chi_{_P}({q})\gamma_\mu]
=\sqrt{N_c}\int\frac{d^3{\vec q}}{(2\pi)^3}\mathrm{Tr}[\varphi_{_P}({\vec q})\gamma_\mu],
\end{equation}
where $\chi_{_P}$ is BS wave function of the meson without instantaneous approach, $N_c=3$ is the color number. The decay constant expressions with different wave functions are presented in the last column of Table \ref{Mass}.

\section{Results and Discussion}

The input parameters are fixed by fitting the mass spectra of charmonium and bottomonium  with experimental data using wave function representation, $\varphi_{1}$, and were not readjusted except $V_0$ when other wave function representations were adopted. We take $m_c = 1.62$ GeV, $m_b = 4.96$ GeV, $\alpha = 0.06$ GeV. For charmonium, $\lambda = 0.21$ GeV$^2$, $\Lambda_{QCD}=0.27$ GeV, while for bottomonium, $\lambda = 0.2$ GeV$^2$, $\Lambda_{QCD}=0.21$ GeV.

\begin{table}[ht]
\begin{center}
\caption{Charmonium mass spectrum (in unit of MeV) and decay constant formulas with different wave function (WF) choices, EX is the central value of experimental data \cite{PDG}.}\label{Mass}
\setlength{\tabcolsep}{3mm}{
\begin{tabular}{c c c c c c c}
\hline
 & $M _{{\psi(1S)}}$ &  $M _{{\psi(2S)}}$  & $M _{{\psi(1D)}}$  &  $M _{{\psi(3S)}}$  & $M _{{\psi(2D)}}$ & $F_V$ \\
\hline
WF\textbackslash EX &3096.90 &3686.1 &3773.7 &4039 &4191 &\\
$\varphi_{\text{1}}$ & 3090.8 & 3683.0 & 3773.6 & 4051.4 & 4105.5 & $\scriptstyle  4\sqrt{3} \int \frac{\mathrm{d}\vec{q}}{(2\pi)^{3}} \left[ f_{5} - \frac{\vec{q}^{2}}{3M^{2}} f_{3} \right]$  \\
$\varphi_{\text{2}}$ & 3046.0 & 3667.3 & 3773.3 & 4052.2 & 4130.6 & $\scriptstyle 4\sqrt{3} \int \frac{\mathrm{d}\vec{q}}{(2\pi)^{3}} \left[ f_{5} - \frac{\vec{q}^{2}}{3M^{2}} f_{3} \right]$ \\
$\varphi_{\text{3}}$ & 3026.0 & 3633.5 & 3773.7 & 4013.0 & 4133.9 & $\scriptstyle 4\sqrt{3} \int \frac{\mathrm{d}\vec{q}}{(2\pi)^{3}} \left[ f_{5} - \frac{\vec{q}^{2}}{3M^{2}} f_{3} \right]$ \\
$\varphi_{\text{4}}$ & $-$    & $-$    & 3773.4 & $-$    & 4186.7 & $\scriptstyle  -4\sqrt{3} \int \frac{\mathrm{d}\vec{q}}{(2\pi)^{3}}  \frac{\vec{q}^{2}}{3M^{2}} f_{3}$ \\
\hline
\end{tabular}}
\end{center}
\end{table}

The calculated charmonium masses and experimental data \cite{PDG} are shown in Table \ref{Mass}. It can be seen that most of the theoretical results obtained using different wave function representations agree with the experimental data.
For the Choice 4 ($\varphi_4$), only the mass spectrum dominated by $D$-wave states is obtained. This is because the input wave function $\varphi_{4}$ lacks a separate $S$-wave term, the $f_3$ and $f_4$ terms are predominantly $D$-wave, and the extracted $S$-wave is subordinate. In the other three wave function representations, independent $S$-wave and $D$-wave terms are included, allowing the solutions to simultaneously provide solutions dominated by $S$-wave or $D$-wave. It should be noted that we have only provided the wave function representations. Whether the eigenvalue is $S$-wave or $D$-wave dominant is not manually adjusted; we can only determine the nature of the state based on the wave function solution and the mass eigenvalue.

\begin{table}[ht]
\begin{center}
\caption{Bottomonium mass spectrum (MeV) with different wave function (WF) choices, EX is the central value of experimental data \cite{PDG}.}\label{B.Mass}
\setlength{\tabcolsep}{3mm}{
\begin{tabular}{c c c c c c }
\hline
 & $M _{{\Upsilon(1S)}}$ &  $M _{{\Upsilon(2S)}}$  & $M _{{\Upsilon(1D)}}$  &  $M _{{\Upsilon(3S)}}$  & $M _{{\Upsilon(2D)}}$  \\
\hline
WF\textbackslash EX  & 9460.3 & 10023.26 &$-$ & 10355.2 & $-$ \\
$\varphi_{\text{1}}$ & 9460.2 & 10021 & 10138 & 10362 & 10436 \\
$\varphi_{\text{2}}$ & 9460.4 & 10031 & 10154 & 10379 & 10454 \\
$\varphi_{\text{3}}$ & 9460.2 & 10025 & 10156 & 10370 & 10458 \\
$\varphi_{\text{4}}$ & $-$    & $-$   & 10137 & $-$   & 11567 \\
\hline
\end{tabular}}
\end{center}
\end{table}

Table \ref{B.Mass} presents the corresponding mass spectrum of bottomonium. In comparison with charmonium, it can be seen that the bottomonium mass agrees perfectly with experimental data \cite{PDG}. This is because bottomonium is significantly heavier, making the instantaneous approximation more suitable. It is almost impossible to determine which choice is better.

\begin{table}[ht]
\begin{center}
\caption{The decay widths $\Gamma$ (keV) of $\psi\to e^+e^-$ using wave functions (1-4).}\label{All result}
\setlength{\tabcolsep}{3mm}{
\begin{tabular}{ c c c c c c}
\hline
  & $\scriptstyle\Gamma _{{\psi(1S)}}$ &  $\scriptstyle\Gamma _{{\psi(2S)}}$  & $\scriptstyle\Gamma _{{\psi(1D)}}$ &  $\scriptstyle\Gamma _{{\psi(3S)}}$  & $\scriptstyle\Gamma _{{\psi(2D)}}$  \\
\hline
$\scriptstyle{\rm WF}$\textbackslash $\scriptstyle{\rm EX}$  & $\scriptstyle5.53\pm0.10$ & $\scriptstyle 2.33\pm0.04$ &$\scriptstyle 0.262\pm0.018$ & $\scriptstyle0.86\pm0.07$  & $\scriptstyle0.48\pm0.22 $  \\
$\scriptstyle\varphi_{1}$  & $\scriptstyle 8.49^{+0.72}_{-0.68}$  & $\scriptstyle3.94^{+0.27}_{-0.24}$ & $\scriptstyle 0.054^{+0.015}_{-0.011} $     & $\scriptstyle 2.57^{+0.14}_{-0.13}$ & $\scriptstyle 0.088^{+0.028}_{-0.020} $ \\
$\scriptstyle\varphi_{2}$  & $\scriptstyle 8.91^{+0.74}_{-0.70}$  & $\scriptstyle 3.38^{+0.10}_{-0.10}$ & $\scriptstyle 0.231^{+0.091}_{-0.067} $     & $\scriptstyle 1.22^{+0.18}_{-0.16}$ & $\scriptstyle 0.986^{+0.240}_{-0.221} $ \\
$\scriptstyle\varphi_{3}$  & $\scriptstyle 12.8^{+1.5}_{-1.3}$    & $\scriptstyle 6.51^{+0.55}_{-0.53}$ & $\scriptstyle 0.378^{+0.151}_{-0.110} $     & $\scriptstyle  3.69^{+0.13}_{-0.16}$ & $\scriptstyle  1.31^{+0.42}_{-0.32}$ \\
$\scriptstyle\varphi_{4}$  & $-$ & $-$ & $\scriptstyle 1.54^{+0.07}_{-0.08}$ & $-$ & $\scriptstyle 0.400^{+0.000}_{-0.003} $ \\
\hline
\end{tabular}}
\end{center}
\end{table}

Using the decay constant formulas given in Table \ref{Mass}, we  calculate the annihilation of quarkonia into dileptons, with the results listed in Tables \ref{All result} and \ref{All result for bbbar}. As shown in Table \ref{All result}, although it is difficult to determine the best wave function representation solely by the mass spectrum, the partial widths to dileptons provide a clear answer. If only considering states dominated by $S$-waves, selecting wave functions $\varphi_{\text{1}}$ and $\varphi_{\text{2}}$ can yield results that are consistent with experiments. However, when also comprehensively considering states dominated by $D$-waves, only the $\varphi_{\text{2}}$ produces the results, $\Gamma({\psi(3770) \to e^+e^-})= 0.231^{+0.091}_{-0.067} $ keV and $\Gamma({\psi(4160) \to e^+e^-})= 0.986^{+0.240}_{-0.221} $ keV,  that match the PDG data $\Gamma^{\rm EX}_{\psi(3770)}= 0.262\pm0.018 $ keV and $\Gamma^{\rm EX}_{\psi(4160)}= 0.48\pm0.22 $ keV \cite{PDG}. Our result of $\psi(4160)$ is in good agreement with the experimental value of $0.83\pm0.07$ keV in Ref. \cite{seth}. We note that the central value of $\Gamma^{\rm EX}_{\psi(4160)\to\mu^+\mu^-}= 2.45\pm1.24\pm0.94 $ keV in recent experiment is quite large \cite{PDG}, indicating the dilepton process of $\psi(4160)$ still requires more precise experimental investigation. Our theoretical error is obtained by varying all parameters by $\pm5\%$ arbitrarily.

\begin{table}[ht]
\begin{center}
\caption{The decay widths $\Gamma$ (keV) of $\Upsilon\to e^+e^-$ using wave functions (1-4).}\label{All result for bbbar}
\setlength{\tabcolsep}{3pt}{
\begin{tabular}{c c c c c c }
\hline
   & $\scriptstyle\Gamma _{{\Upsilon(1S)}}$ &  $\scriptstyle\Gamma _{{\Upsilon(2S)}}$  & $\scriptstyle\Gamma _{{\Upsilon(1D)}}{\times10^{-3}}$ & $\scriptstyle\Gamma _{{\Upsilon(3S)}}$ &  $\scriptstyle\Gamma _{{\Upsilon(2D)}}{\times10^{-3}}$ \\
\hline
$\scriptstyle{\rm WF}$\textbackslash $\scriptstyle{\rm EX}$   & $\scriptstyle 1.340\pm0.018$ & $\scriptstyle 0.612\pm0.011$ &- & $\scriptstyle0.443\pm0.008$  & -  \\
$\scriptstyle\varphi_{1}$   & $\scriptstyle 1.29^{+0.09}_{-0.08}$ & $\scriptstyle 0.630^{+0.048}_{-0.046}$ &  $\scriptstyle 0.748^{+0.232}_{-0.150}$  &  $\scriptstyle 0.447^{+0.034}_{-0.032}$ &  $\scriptstyle 1.18^{+0.36}_{-0.24}$ \\
$\scriptstyle\varphi_{2}$  & $\scriptstyle 1.32^{+0.10}_{-0.06}$ & $\scriptstyle 0.610^{+0.043}_{-0.042}$ &  $\scriptstyle 2.29^{+0.86}_{-0.69}$  &  $\scriptstyle 0.412^{+0.028}_{-0.027}$ &  $\scriptstyle 10.5^{+4.2}_{-3.1}$ \\
$\scriptstyle\varphi_{3}$ & $\scriptstyle 1.56^{+0.11}_{-0.14}$ & $\scriptstyle 0.788^{+0.074}_{-0.068}$ &  $\scriptstyle 2.98^{+1.24}_{-0.86}$  &  $\scriptstyle 0.564^{+0.051}_{-0.047}$ &  $\scriptstyle 12.5^{+4.9}_{-3.5}$ \\
$\scriptstyle\varphi_{4}$ & $ -$ & $-$ &  $\scriptstyle 2330^{+80}_{-80}$  & $-$ &  $\scriptstyle 195^{+7}_{-8}$ \\
\hline
\end{tabular}}
\end{center}
\end{table}

As shown in Table \ref{All result for bbbar} for the annihilation decay results of bottomonium, if only the $S$-wave dominated states are considered, wave functions $\varphi_{\text{1}}$ and $\varphi_{\text{2}}$ can all yield results consistent well with experimental data \cite{PDG}. However, we believe that the same wave function representation should be adopted for both charmonium and bottomonium. Therefore, we propose selecting wave function $\varphi_{\text{2}}$, for the $D$-wave dominated states, we obtain $\Gamma ({\Upsilon(1D)} \to e^+e^-)=  2.29^{+0.86}_{-0.69}$ eV and $\Gamma ({\Upsilon(2D)} \to e^+e^-)= 10.5^{+4.2}_{-3.1}$ eV.

For comparison, we present in Table \ref{theory1} the results calculated using wave function $\varphi_{\text{2}}$ as well as results from other theories. As can be seen, we have provided the largest decay width, especially the $2D$ result, which is several times larger than other theoretical results. Apart from Refs. \cite{1983} and \cite{zhong}, which account for $S-D$ mixing by incorporating tensor force and coupled-channel effects respectively, other theoretical calculations have neglected the mixing effect. From the results, even when mixing effects are considered, the predictions in Refs. \cite{1983} and \cite{zhong} remain very close to other theoretical values, indicating that the mixing effect in these two papers is minimal. This observation is inconsistent with our findings, particularly regarding the $2D$ results.

\begin{table}[ht]
\begin{center}
\caption{Dilepton decay widths (eV), ours are obtained using the wave function $\varphi_{2}$}\label{theory1}
\setlength{\tabcolsep}{3mm}{
\begin{tabular}{c c c cc c c c c c}
\hline
 & ours &\cite{1983}&\cite{Gonzalez}&\cite{Badalian2}& \cite{godfrey} &  \cite{Segovia}  & \cite{xliu} &\cite{kher}&\cite{zhong}  \\
\hline
$\Gamma ({\Upsilon(1D)} \to e^+e^-)$ &$2.29^{+0.86}_{-0.69}$ & 1.5&0.37 &0.62& 1.38 &1.4 &1.88 &1.65&1.08 \\
$\Gamma ({\Upsilon(2D)} \to e^+e^-)$ &$10.5^{+4.2}_{-3.1}$ & 2.7 & 0.58&1.08& 1.99 &2.5 &2.81 &2.42&2.13 \\
\hline
\end{tabular}}
\end{center}
\end{table}

Next, we discuss the mixing problem. Unlike the commonly adopted approach, we do not separately solve for the $S$-wave and $D$-wave wave functions and artificially mix them using Eq. (\ref{mix}), with the undetermined mixing angle fitted to experimental values. Instead, we present a wave function representation for the $1^{--}$ state that simultaneously contains $S$-wave, $P$-wave, and $D$-wave components, where the relative proportions of different waves are determined by the dynamical Salpeter equation satisfied by the wave function.

\begin{figure}[htbp]
    \centering
    \begin{subfigure}[b]{0.3\textwidth}
        \centering
        \includegraphics[width=\textwidth]{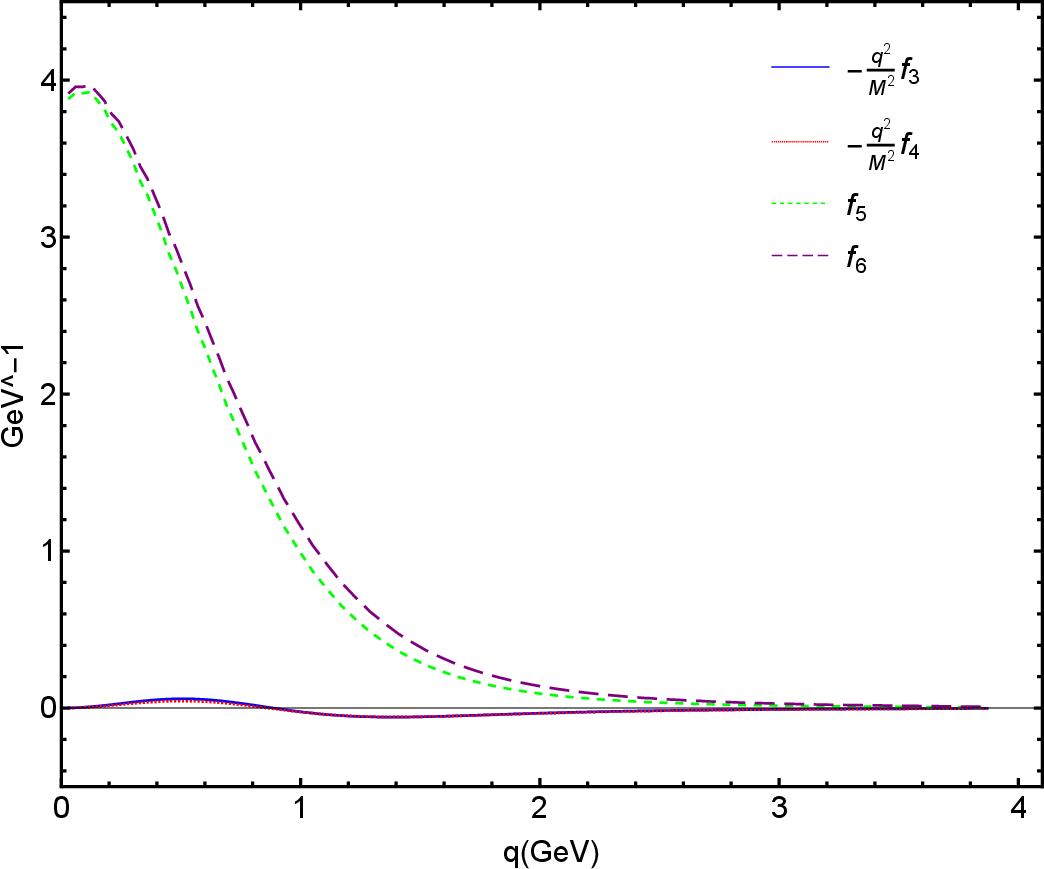}
        \caption{$\psi(1S)$}
        \end{subfigure}
        \hfill
        \begin{subfigure}[b]{0.3\textwidth}
        \centering
        \includegraphics[width=\textwidth]{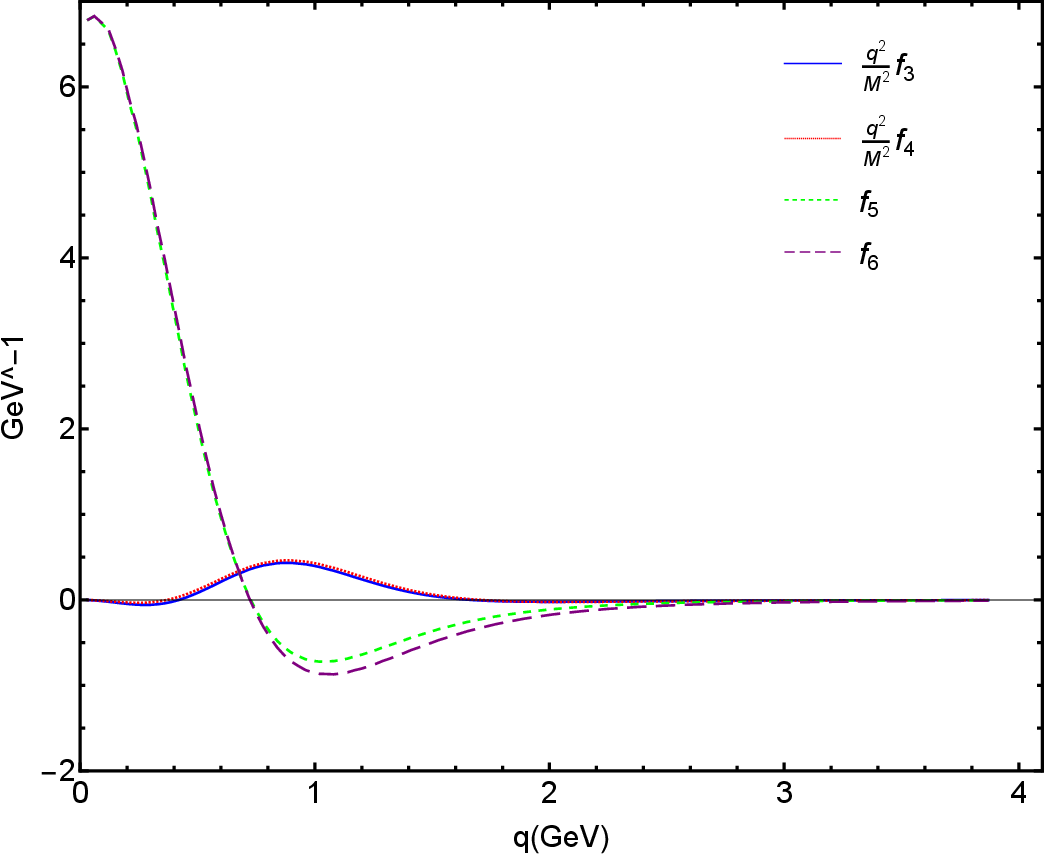}
        \caption{$\psi(2S)$}
        \end{subfigure}
        \hfill
    \begin{subfigure}[b]{0.3\textwidth}
        \centering
        \includegraphics[width=\textwidth]{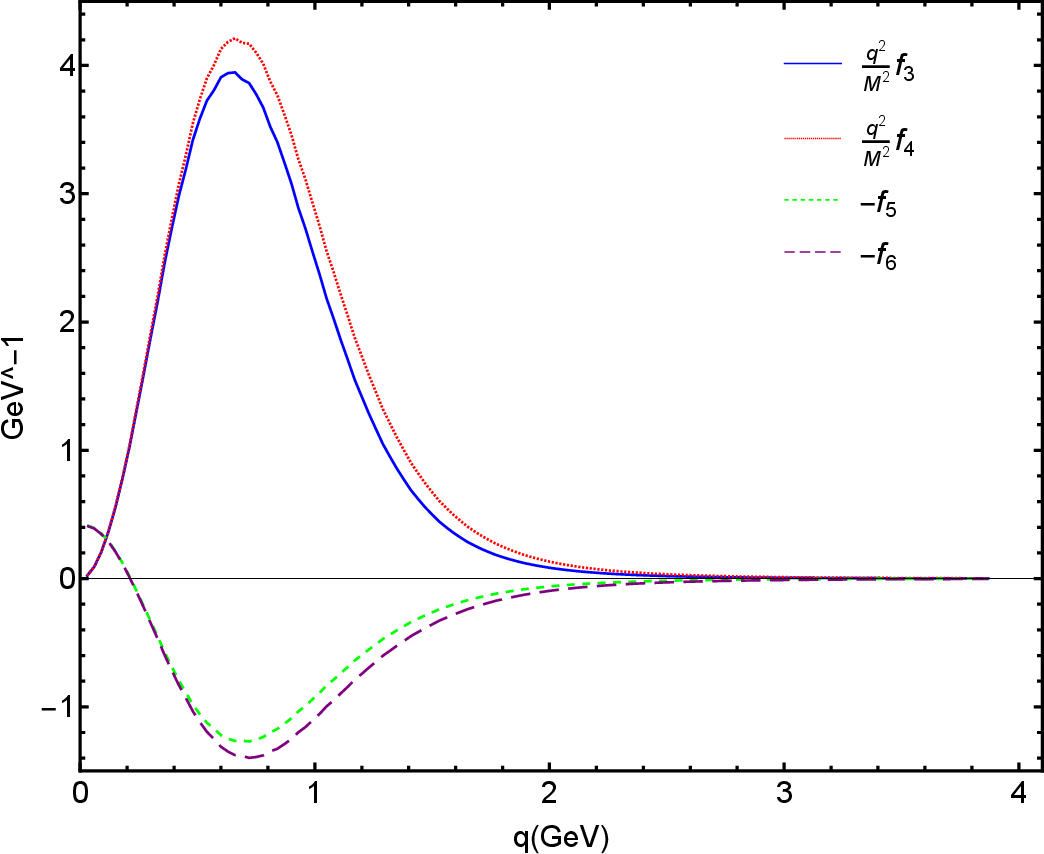}
        \caption{$\psi(1D)$}
    \end{subfigure}
 \vspace{0.5cm}
    \begin{subfigure}[b]{0.3\textwidth}
        \centering
        \includegraphics[width=\textwidth]{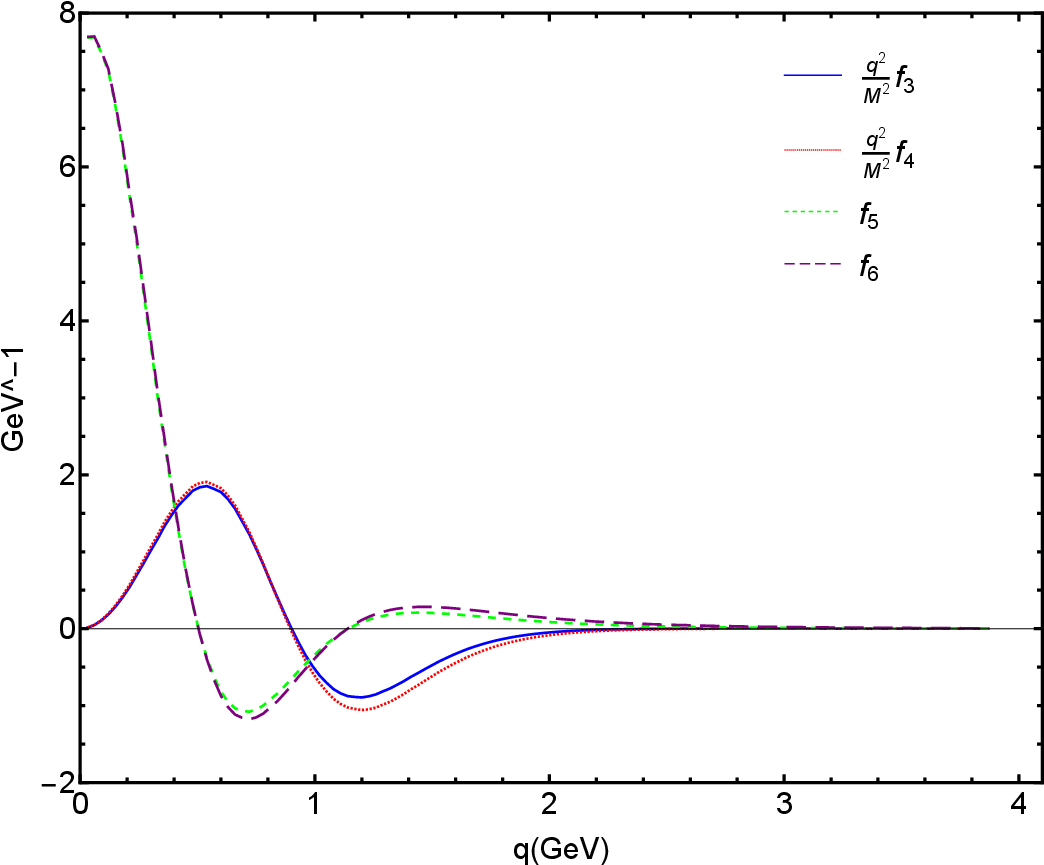}
        \caption{$\psi(3S)$}
    \end{subfigure}
    \begin{subfigure}[b]{0.3\textwidth}
        \centering
        \includegraphics[width=\textwidth]{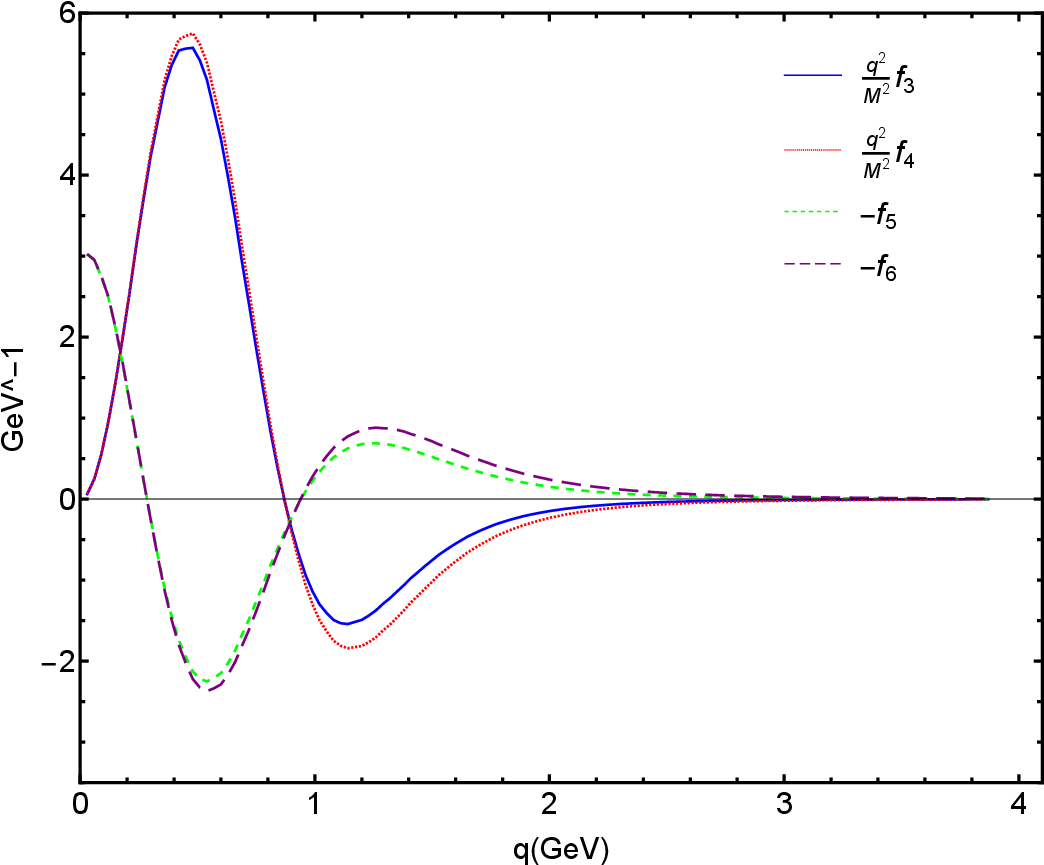}
        \caption{$\psi(2D)$}
    \end{subfigure}
    \caption{Radial wave functions of the wave function $\varphi_{2}$ for charmonium, where $q\equiv |\vec{q}|$.}
    \label{figs}
\end{figure}

We illustrate this using only charmonium wave function $\varphi_{2}$ as an example. The independent radial wave functions corresponding to the first five eigen states are plotted in Fig. \ref{figs}. In Fig. \ref{figs}(a), the $S$-wave radial wave functions $f_5$ and $f_6$ dominate and have no nodes, indicating that this state is the $1S$ state $\psi(1S)$. Its wave function contains a very small admixture of $D$-wave from $f_3$ and $f_4$, which manifests as a $2D$-wave. In Fig. \ref{figs}(b), the dominant $S$-wave has one node, identifying it as the $2S$ state $\psi(2S)$, while the mixed $D$-wave component overall exhibits the behavior of a $1D$-wave. In Fig. \ref{figs}(c), $f_3$ and $f_4$ dominate and show no nodes, thus corresponding to the $1D$ state $\psi(1D)$ mixed with a $2S$-wave. In Fig. \ref{figs}(d), the $3S$-wave is dominant and mixed with a $2D$-wave, so it is the $\psi(3S)$; whereas the fifth state in Fig. \ref{figs}(e) is predominantly a $2D$-wave mixed with a $3S$-wave, corresponding to $\psi(2D)$ state.

In addition to the $S$-wave and $D$-wave, the $1^{--}$ wave function in our method also contains a $P$-wave component. Rather than plotting its radial wave function, we instead present the proportions of the $S$-, $P$-, and $D$-wave components in each state based on the normalized wave function formula.
For example, the normalization of the wave function in Eq. (\ref{1-}) is given in the first row of Table \ref{rela} as $\int \frac{d\vec{q}}{(2\pi)^3} \frac{4Mw_1}{3m_1}\left[ 3f_5f_6+\frac{q_\perp^2}{M^2}\left(\frac{q_\perp^2}{M^2}f_3f_4+f_3f_6 +f_4f_5 \right) \right]=1$$\equiv(N_S+N_P+N_D)^2$, which corresponds to $(S+P+D)^2$. When only the $S$-wave is present, the left-hand side of the formula becomes
$\int \frac{d\vec{q}}{(2\pi)^3} \frac{4Mm_1}{w_1}\left(f_5+\frac{q_\perp^2}{3M^2}f_3 \right)\left(f_6+\frac{q_\perp^2}{3M^2}f_4 \right)$, corresponding to $S^2$; when only the $D$-wave is present, the left-hand side becomes $\int \frac{d\vec{q}}{(2\pi)^3} \frac{8m_1q_\perp^4f_3f_4}{9w_1M^3}$, corresponding to $D^2$. Thus, using these three formulas, we can solve for the proportions of the $S$-wave, $P$-wave, and $D$-wave components.

{We need to point out that due to differences in both the representations and the normalizations, the $S:P:D$ ratios given by the different wave functions in Eq. (\ref{wave}) will differ. However, we do not intend to present all cases; we only provide the results for the wave function $\varphi_{2}$ in Table \ref{SPD}. As can be seen from the Table \ref{SPD},} in the $1S$, $2S$, and $3S$ states, the $S$-wave dominates and provides the non-relativistic contribution, the $P$-wave and $D$-wave  supply the relativistic correction. In the $1D$ and $2D$ states, the $D$-wave is dominant and provides the non-relativistic contribution, the $S$-wave and $P$-wave contribute the relativistic corrections. Comparing the ratios of $S:P:D$ reveals that the relativistic corrections in bottomonium are much smaller than those in charmonium. Furthermore, it can also be seen from the $S:D$ ratios that the mixing angles in bottomonium are significantly smaller than those in charmonium.

\begin{table}[ht]
\begin{center}
\caption{Partial wave ratios $S:P:D$ in the wave function $\varphi_{2}$ for heavy quarkonium}\label{SPD}
\setlength{\tabcolsep}{3mm}{
\begin{tabular}{c c c c c c }
\hline
 & $1S$ &  $2S$  & $1D$  &  $3S$  & $2D$  \\
\hline
$\psi$ & $\scriptstyle 1~:~0.0639~:~0.0281$&$\scriptstyle 1~:~0.180~:~0.117$ &$\scriptstyle 0.139~:~0.221~:~1$ &$\scriptstyle 1~:~0.252~:~0.483 $& $\scriptstyle 0.495~:~0.339~:~1$ \\
$\Upsilon$ & $\scriptstyle 1~:~0.0257~:~0.0137$&$\scriptstyle 1~:~0.0339~:~0.0208$ &$\scriptstyle 0.0310~:~0.0499~:~1$ &$\scriptstyle 1~:~0.0617~:~0.0816 $& $\scriptstyle 0.0953~:~0.0726~:~1$ \\
\hline
\end{tabular}}
\end{center}
\end{table}

Since the wave functions in Eq. (\ref{wave}) contain $S$-wave, $P$-wave, and $D$-wave, we cannot directly define a mixing angle. However, common approaches in the literature typically mix only $S$-wave and $D$-wave components and determine the mixing angle by comparing with experimental data. To facilitate comparison with these existing theoretical results, we neglect the $P$-wave in our method and retain only the $S$-wave and $D$-wave components, thereby allowing us to define a mixing angle in the conventional sense and make comparisons with other results.

{The original wave functions with $S+P+D$ components are normalized, i.e.,
$(N_S+N_P+N_D)^2=1$. To compute the mixing angle, we discard the $P$-wave contribution and re-normalize. Therefore, for an $S$-wave-dominated $2S$ state, its normalization is
$(N_{S1}+N_{D1})^2=1$, and the mixing angle $\theta$ is defined via the first relation in Eq. (\ref{mix}), giving $\cos\theta=N_{S1}$ and $\sin\theta=N_{D1}$, i.e., $\theta_{2S}=\arctan(N_{D1}/N_{S1})$. For a $D$-wave-dominated $1D$ state, its normalization is $(N_{S2}+N_{D2})^2=1$, and the second relation in Eq. (\ref{mix}) applies, yielding
$\sin\theta=N_{S2}$ and $\cos\theta=N_{D2}$, i.e., $\theta_{1D}=\arctan(N_{S2}/N_{D2})$.}
So, the mixing angles for the $2S$ and $1D$ states merely borrow the two formulas from Eq. (\ref{mix}) for their definitions, but are solved independently. In principle, these two angles are unrelated; thus we obtain two mixing angles. The results are presented in Table \ref{mxingangle}. As can be seen, the mixing angles we obtained for $2S$ and $1D$ states, as well as for $3S$ and $2D$ states, are generally different. For example, adopting $\varphi_{2}$, $\theta_{2S}=(6.67^{+1.71}_{-1.50})^{\circ}$ is obtained for $\psi(2S)$, while for $\psi(1D)$ it is $\theta_{1D}=(7.92^{+1.70}_{-1.43})^{\circ}$. The latter is quite close to the ones from the simple $2S-1D$ mixing scheme: $10^{\circ}$ \cite{Ding:1991vu}, $12^{\circ}$
\cite{smallangle1}, and $(12\pm 2)^{\circ}$ \cite{smallangle2}.
For the $2D$ state $\psi(4160)$, as a $3S-2D$ mixing state, $\theta_{2D}=(26.3^{+3.3}_{-3.4})^\circ$ is obtained, which confirms the large mixing angle of
$35^{\circ}$ in Ref. \cite{chao661}, $34.8^{\circ}$ in Ref. \cite{Badalian}, and $21.2^{\circ}$ in Ref. \cite{zzhao}.

\begin{table}[ht]
\begin{center}
\caption{The $S-D$ mixing angle $\theta$ ($\, ^\circ$) of charmonium and bottomonium with neglected the $P$-wave using wave functions (1-4).}\label{mxingangle}
\setlength{\tabcolsep}{3mm}{
\begin{tabular}{c c c c c c c c c}
\hline
 $\scriptstyle{\rm WF}$ & $\scriptstyle\theta _ {{\psi(2S)}}$ & $\scriptstyle\theta _ {{\psi(1D)}}$ & $\scriptstyle\theta _ {{\psi(3S)}}$ & $\scriptstyle\theta _ {{\psi(2D)}}$ & $\scriptstyle \theta _ {{\Upsilon(2S)}}$ & $\scriptstyle \theta _ {{\Upsilon(1D)}}$& $\scriptstyle \theta _ {{\Upsilon(3S)}}$ & $\scriptstyle \theta _ {{\Upsilon(2D)}}$   \\
\hline
$\scriptstyle\varphi_{1}$ & $\scriptstyle 3.66^{+0.06}_{-0.15} $ & $\scriptstyle 3.65^{+0.27}_{-0.25} $ &$\scriptstyle 4.02^{+0.23}_{-0.22} $ &$\scriptstyle 4.01^{+0.16}_{-0.21} $
 & $\scriptstyle 1.44^{+0.11}_{-0.10}$ & $\scriptstyle 1.16^{+0.10}_{-0.10}$ & $\scriptstyle 1.56^{+0.12}_{-0.12}$      & $\scriptstyle 1.41^{+0.12}_{-0.12}$  \\
$\scriptstyle\varphi_{2}$ & $\scriptstyle 6.67^{+1.71}_{-1.50}  $ & $\scriptstyle 7.92^{+1.70}_{-1.43}  $ & $\scriptstyle 25.8^{+3.0}_{-3.7}  $& $\scriptstyle 26.3^{+3.3}_{-3.4}  $
& $\scriptstyle 1.19^{+0.22}_{-0.04}$ & $\scriptstyle 1.78^{+0.32}_{-0.25}$   & $\scriptstyle 4.67^{+0.85}_{-0.65}$    & $\scriptstyle 5.44^{+1.10}_{-0.76}$ \\
$\scriptstyle\varphi_{3}$ & $\scriptstyle 4.45^{+0.77}_{-0.80}  $ & $\scriptstyle 5.97^{+0.73}_{-0.69}  $ & $\scriptstyle 13.7^{+1.3}_{-1.2}  $ & $\scriptstyle 14.6^{+1.0}_{-1.2}  $
& $\scriptstyle 1.12^{+0.19}_{-0.18}$ & $\scriptstyle 1.68^{+0.26}_{-0.20}$   & $\scriptstyle 3.97^{+0.59}_{-0.58}$    & $\scriptstyle 4.61^{+0.62}_{-0.54}$  \\
$\scriptstyle\varphi_{4}$ & $\scriptstyle -$  & $\scriptstyle 35.3\pm{0}$ & $\scriptstyle -$  & $\scriptstyle 35.3\pm{0}$ & $\scriptstyle -$  & $\scriptstyle 35.3\pm{0}$         & $\scriptstyle -$      & $\scriptstyle 35.3\pm{0}$  \\
\hline
\end{tabular}}
\end{center}
\end{table}

For bottomonium, the mixing angles we obtained using $\varphi_{2}$ are ($\theta_{\Upsilon(1D)}={(1.78^{+0.32}_{-0.25}})^{\circ}$, $\theta_{\Upsilon(2D)}=({5.44^{+1.10}_{-0.76}})^{\circ}$), which are significantly larger than the results from Ref. \cite{zou2} using the coupled-channel approach, (0.02$^{\circ}$, 0.27$^{\circ}$). Ref. \cite{Moremix3} also chose the coupled-channel method, if we apply their data and only account for $2S-1D$ and $3S-2D$ mixings while ignoring other mixings, their mixing angles are (2.63$^{\circ}$, 5.60$^{\circ}$), which align well with ours. In addition to small mixing angles, Ref. \cite{zzhao} reports large mixing angles,  (9$^{\circ}$, 12.5$^{\circ}$). Due to the absence of experimental data for the $1D$ and $2D$ states, they determined the mixing angles by fitting only the dilepton decays of the $2S$ and $3S$ states. However, as known that the $S$-wave component in the $1D$ state contributes the majority of the dilepton partial width, while the $D$-wave component has only a minor effect on the partial width of the $2S$ state. Considering other influences such as relativistic and QCD corrections, extracting mixing angles solely by fitting the $2S$ and $3S$ states makes it difficult to control the errors.

\section{Conclusion}

This paper systematically investigates four different relativistic wave function representations by solving the instantaneous BS equation, with a focus on exploring the $S-D$ mixing in vector charmonium and bottomonium states.

We find that the choice of wave function representation is crucial. Among the four distinct representations we have constructed ($\varphi_1$, $\varphi_2$, $\varphi_3$, and $\varphi_4$), the $\varphi_2$ representation provides the most consistent description of the experimentally observed mass spectra and dileptonic decay widths. This representation not only accurately reproduces the properties of $S$-wave dominant states (such as $\psi(1S-3S)$, $\Upsilon(1S-3S))$, but also successfully predicts the decay behaviors of $D$-wave dominant states (such as $\psi(3770)$ and $\psi(4160)$), with theoretical calculations in good agreement with experimental data. Based on this, we predict the decay widths of the yet-unobserved $\Upsilon(1D)$ and $\Upsilon(2D)$ states.

$S-P-D$ mixing is a universal phenomenon: Our analysis shows that the wave functions of vector mesons are not simple $S-D$ mixtures but contain significant $P$-wave components, forming $S-P-D$ mixing. For example, in the $\psi(2S)$ state, the proportions of $S$, $P$, and $D$ waves are approximately $1:0.18:0.12$; in the $\psi(3770)$, the ratio is about $0.14:0.22:1$. Such mixing is naturally determined by the dynamics of the Salpeter equation, reflecting the inherent requirements of relativistic effects.

Based on calculations using the wave function $\varphi_2$, we obtain more reasonable mixing angles. For charmonium, the mixing angle of $\psi(3770)$ is $(7.92^{+1.70}_{-1.43})^\circ$, and that of $\psi(4160)$ is $(26.3^{+3.3}_{-3.4})^\circ$. The mixing angles in bottomonium are much smaller than those in charmonium, such as $(1.78^{+0.32}_{-0.25})^\circ$ for $\Upsilon(1D)$ and $(5.44^{+1.10}_{-0.76})^\circ$ for $\Upsilon(2D)$, owing to the heavier mass of the bottom quark, which reduces relativistic effects.

This paper predicts the dileptonic decay widths of the bottomonium states $\Upsilon(1D)$ and $\Upsilon(2D)$ to be $(2.29^{+0.86}_{-0.69})$ eV and $(10.5^{+4.2}_{-3.1})$ eV, respectively. These values are larger than existing theoretical results, and future experimental measurements will help verify the reliability of this model.

In summary, by optimizing the wave function representation, this study achieves an accurate description of $S-P-D$ mixing in vector quarkonia within the Salpeter equation framework, emphasizing the importance of relativistic corrections and multi-wave mixing. The results provide valuable theoretical references for future experimental studies of related particles.

{\bf Acknowledgments}
This work was supported by the National Natural Science Foundation of China (NSFC) under the Grant No. 12575097, by the Natural Science Foundation of Guangxi Autonomous Region with Grant No. 2022GXNSFDA035068, and by the NSFC under the Grant No. 12075073.
Q. Li was supported by the National Key R\&D Program of China\,(2022YFA1604803) and the Natural Science Basic Research Program of Shaanxi\,(No.\,2025JC-YBMS-020). T. Wang was supported by the NSFC under the Grant No. 12375085 and the Fundamental Research Funds for the Central Universities (2023FRFK06009).

\end{document}